\documentstyle[preprint,aps,prd,psfig,here]{revtex}
\tightenlines
\def\nn{\nonumber}
\def\beq{\begin{equation}}
\def\eeq{\end{equation}}
\def\bea{\begin{eqnarray}}
\def\eea{\end{eqnarray}}
\def\nn{\nonumber}

\def\Fk{F_K}
\def\Ms{M_{\sigma}}
\def\Fp{F_\pi}
\def\gb{g_2}
\def\ga{g_1}
\def\Ms{M_{\sigma}}
\def\dU{\partial U}
\def\dUd{\partial U^\dagger}

\def\DS{DS}
\def\dK{\partial K^0}
\def\dKb{\partial {\bar K^0}}
\def\dpi{\partial {\pi^0}}
\def\K{K^0}
\def\Kb{{\bar K^0}}
\def\Mp{M_{\pi}}
\def\Mk{M_K}
\def\M{M_{\sigma}}
\def\Kpipi{K \rightarrow \pi^0 \pi^0}
\def\Bb{\bar B}
\def\Bzb{\bar B_0}
\def\epe{{\epsilon' / \epsilon}}
\def\lrpartial{\stackrel{\leftrightarrow}{\partial}}

\def\Kpipi{K_s \rightarrow \pi^0 \pi^0}
\def\Kpm{K_s \rightarrow \pi^+ \pi^-}
\def\Kpp{K^+ \rightarrow \pi^0 \pi^+}
\def\Kppg{K \rightarrow \pi \pi}
\def\hyf{\mbox{-}}
\begin{document}

\title{Penguin diagrams in the $\Delta I=1/2$ rule and $\epsilon'/\epsilon$ with $\sigma$ models}

\author{M. Harada $^\dag$
}
\address{Department of Physics, Nagoya University,\\ Nagoya, 464-8602, Japan }

\author{Y. Keum $^\ast$
}
\address{Theory group, KEK \\ Tsukuba, Ibaraki 305-0801, Japan}

\author{
Y. Kiyo $^\clubsuit$
, T. Morozumi $^\diamondsuit$
, T. Onogi $^\heartsuit$ 
, N. Yamada $^\spadesuit$
}

\address{ Department of Physics, Hiroshima University \\
1-3-1 Kagamiyama, Higashi Hiroshima  -  739-8526, Japan}

\date{\today}

\preprint{HUPD-9913, ~KEK-TH-643, ~DPNU-99-27}

\maketitle

\begin{abstract}

    We study the correlation between $\epe$ and $\Delta I=1/2$ rule 
    in the framework of non-linear $\sigma$ model including scalar mesons.
    Using this model we estimate the chiral corrections by resumming 
    to all orders in Chiral Perturbation Theory, the contribution 
    of a class of diagrams within the factorization approximation. 
    With these matrix elements and changing the scalar meson mass,  
    we find that there is correlation between $\epe$ and $\Delta I=1/2$ amplitude.
    However, it is difficult to explain both $\epe$ and $\Delta I=1/2$ amplitude 
    simultaneously. In order to be compatible with $\epe$, typically, about half 
    of $\Delta I=1/2$ amplitude can be explained at most. Our result suggests there
    may be substantial non-factorizable contribution to CP conserving $K \rightarrow \pi \pi$ 
    amplitudes.

\end{abstract}

\vspace*{3cm}

\hrulefill

\begin{picture}(500,54)
       \put(0,40)
{$\dag$ E-mail address: harada@eken.phys.nagoya-u.ac.jp}
       \put(0,28)
{$\ast$ E-mail address: ccthmail.kek.jp}
       \put(0,16)
{$\clubsuit$ E-mail address: kiyo@theo3.phys.sci.hiroshima-u.ac.jp}
       \put(0,04)
{$\diamondsuit$ E-mail address: morozumi@theo.phys.sci.hiroshima-u.ac.jp}
       \put(0,-8)
{$\heartsuit$ E-mail address: onogi@theo.phys.sci.hiroshima-u.ac.jp}
       \put(0,-20)
{$\spadesuit$ E-mail address: yamada@theo3.phys.sci.hiroshima-u.ac.jp }
\end{picture}

%
\def\Kpipi{K_s \rightarrow \pi^0 \pi^0}
\def\Kpm{K_s \rightarrow \pi^+ \pi^-}
\def\Kpp{K^+ \rightarrow \pi^0 \pi^+}
\newpage
\section{\bf Introduction}

According to recent measurements of direct CP violation in
$\Kppg$ decays, $\epe$ is  $ O(10^{-3})$ \cite{KTeV,NA31}.
Theoretical prediction ranges around $10^{-4} \sim 10^{-3}$.
It strongly depends on the hadronic matrix elements of QCD 
\cite{VZS} and EW penguin operators \cite{BW,FL,DDG}.

An interesting possibility is suggested as an explanation
for large $\epe=O(10^{-3})$ in the standard model \cite{KNS}.
The authors argue that the mechanism which enhances
$\Delta I=1/2$ amplitude may also enhance $\epe$ and it
naturally leads to the measured values.  
The enhancement comes from the Feynman diagram which includes 
the scalar $\sigma$ meson as intermediate state.
The mechanism was found in the framework of linear 
$\sigma$ model \cite{SHA,MLS}. Though qualitative picture 
of the linear $\sigma$ model may be correct, for quantitative 
analysis, some improvement can be made. In the linear $\sigma$ model 
employed in Ref.\cite{MLS}, $\sigma$  meson mass can be written 
in terms of the physical quantities $F_K, F_\pi, M_K$ and $M_\pi$ 
and its mass is predicted to be about $900$ MeV. 
SU(3) breaking ratio ${m_u /m_s}$ is also determined by the same 
input and numerically it is around ${1/ 30}$. 
The enhancement factor of a QCD penguin operator is written as 
${F_K /(3 F_\pi-2 F_K)}
 =({M_n\sigma}^2-{M_\pi}^2)/({M_\sigma}^2-{M_K}^2) 
  \cdot ({F_K}^2/{F_\pi}^2) \simeq 2.$
These relations and numbers are specific predictions of the 
linear $\sigma$ model. Because the dynamical property of the 
linear $\sigma$ model is not same as that of QCD, these relations 
and numbers may be taken as semi-quantitative \cite{KNS}.

In this paper, we study the matrix elements of the
QCD and EW penguins with the non-linear $\sigma$ model 
including scalar mesons. The model is built with chiral symmetry 
as a guide. It is more general than the linear $\sigma$ model and  
less dependent on dynamical assumption. The cost is that it has more 
parameters. They can be determined with the experimental measured 
quantities, i.e., decay width, mass spectrum, etc.   
Still $\sigma$ meson mass is left as a free parameter because 
the spectroscopy of $\sigma$ meson $f_0(400-1200)$ allows wide 
range for the mass. (See Refs. \cite{HSS,TR,IITITT,MP,JPHS,PDG} 
for scalar meson mass spectroscopy.) 
We study how QCD and EW penguin matrix elements depend on the mass 
of $\sigma$ meson.
%

Here we write a few words on the difference between our approach
and the conventional treatment of the resonances in Chiral Perturbation 
Theory (CHPT). CHPT is a systematic treatment, in the sense of the small 
momentum expansion, and describes low energy processes involved pions 
and kaons well. However higher order terms in CHPT are of great importance 
if threshold of the $\sigma$ meson is near to kaon mass.   
Thus it is very interesting to explore non-linear $\sigma$ model with scalar 
resonances in kaon decays. Correspondence between non-linear $\sigma$ 
model with resonances and CHPT was well discussed in Refs.\cite{EGPR,EKW} 
at order of $p^4$ and their results indicate that the resonance contributions 
dominate the low energy coupling constants in the strong part of $p^4$ chiral 
lagrangian.
In our approach, we compute the full contribution of the $\sigma$ meson to the
factorizable part of QCD and EW penguins using non-linear $\sigma$ model with 
scalar resonances, so that a certain class of higher order terms of $p^n (n>4)$ 
in momentum expansion are included.
We show how these higher order terms contribute to observable 
in kaon decays.

As a phenomenological application, we compute $\epe$ and
$\Delta I=1/2$ amplitude in isospin limit.
We use the Wilson coefficients for four-fermi interaction
at $\mu=0.8 \sim 1.2$ GeV in the next to leading log (NLL)
approximation \cite{BBL,BJL,BBH}.
The four-fermi operator is factorized into
products of color singlet currents (or densities) and they
are identified with those of the $\sigma$ model \cite{BBG}.
The density$\times$density type operators are enhanced for
small strange quark mass by a factor of $(1/ m_s)^2$. Therefore
numerical values for the strange quark mass are important.
As for the strange quark mass, we choose the range which is suggested 
by QCD sum rule \cite{SUMR} and lattice simulations \cite{ONO}. 
We also study the correlation 
between $\epe$ and $\Delta I=1/2$ amplitude. This is done by varying 
the $\sigma$ mass, strange quark mass $m_s$, and factorization scale $\mu$.
By studying the dependence of $\epe$ and $\Delta I=1/2$ amplitude 
on the $\sigma$ meson mass, we search for the range of the mass  
which may reproduce both $\displaystyle{\epe}$ and $\Delta I=1/2$ amplitude.

The paper organized as follows:
In section II, we summarize the outline
of the computation  $\epe$ and $\Delta I=1/2$ amplitude.
In section III,
we derive the matrix elements of penguin operators.
In section IV, numerical results of $\epe$ and $\Delta I=1/2$ are
summarized. In section V, we discuss the implication of our results.
Some useful formulae are collected in appendix.

\section{\bf  $\Delta I=1/2$ rule and ${\epsilon' / \epsilon}$
in the standard Model}
In this section, we summarize our notations and show
outline of computation of $\epe$ and $\Delta I=1/2$ amplitude.
Some details of definitions of isospin amplitudes can be
found in appendix A.
We start with the effective hamiltonian for $\Delta S=1$
non-leptonic decays \cite{BBL},
\bea
H_{eff}= {G_F \over \sqrt{2}} V_{ud} V_{us}^* \sum_{i=1}^{10}
\{z_i + \tau y_i \} Q_i+ h.c,
\eea
where $\tau=-(V_{td}V_{ts}^*) / (V_{ud}V_{us}^*)$.
The isospin amplitudes of $\Kppg$ are defined as
$\langle I|H_{eff}|K^0\rangle =i a_I \exp{i \delta_I}, \quad
\langle I|H_{eff}|\bar{K^0}\rangle =-i {a_I}^* \exp{i \delta_I}
$. $\epsilon'$ is expressed in terms of $a_0$ and $a_2$.
\bea
\epsilon'={1 \over \sqrt{2}} {Re a_2 \over Re a_0} ({Im a_2 \over Re a_2}-
{Im a_0 \over Re a_0}) \exp\{i(\delta_2-\delta_0+ {\pi \over 2}) \}.
\eea
In the factorization
approximation, $a_I$ s are written as,
\bea
&&Re a_I=  {G_F \over \sqrt{2}} Re V_{ud} V_{us}^* \nn \\
&& \times\sum_{i=1}^{5} 
\left[
(z_{2i-1}+{z_{2i}\over{N_c}}) \langle I|Q_{2i-1}|K^0\rangle +
(z_{2i}+{z_{2i-1} \over {N_c}})\langle I|Q_{2i}|K^0\rangle 
\right] \frac{1}{i}, \\
&&Im a_I=-{G_F \over \sqrt{2}} Im V_{td} V_{ts}^* \nn \\
&& \times \sum_{i=1}^{5}
\left[
(y_{2i-1}+{y_{2i}\over{N_c}}) \langle I|Q_{2i-1}|K^0\rangle +
(y_{2i}+{y_{2i-1} \over {N_c}})\langle I|Q_{2i}|K^0 \rangle 
\right] \frac{1}{i},
\eea
where the matrix elements $\langle I|Q_i|K^0\rangle $ is defined in large $N_c$ limit.
As we discuss in detail in the next section, we compute the hadronic matrix
element in large $N_c$ limit,
i.e., we factorize the four-fermi operators
into products of color singlet currents (densities).
The currents (densities) are identified with those
of the chiral lagrangian.
The factorization scale is chosen  at
$0.8$ $-$ $1.2$ GeV, i.e., below charm quark mass $m_c$.
In the factorization approximation,
this choice is mandatory
because
above $m_c$, real part of the Wilson coefficient of
QCD penguin operators
is zero and
it is born below $m_c$
due to the incomplete cancellation of GIM
mechanism. About Wilson coefficients, we use NLL
approximation \cite{BBL,BJL,BBH} and compute them at the
factorization scale. Combining the matrix elements with the
Wilson coefficients, $a_I$ s are given as,
\bea
&&Re a_0= {G_F \over \sqrt{6}} \lambda \frac{X}{i}
 \left[
 2z_2+3z_4-z_1+3 z_6 {Y_6 \over X}
 \right],
 \nn \\
&&Re a_2= {G_F \over \sqrt{3}} \lambda \frac{X}{i}
 \left[z_1+ z_2 \right],  \nn \\
&&Im a_0={-\sqrt{3}G_F \over \sqrt{2}} \frac{X}{i}
 {(A \lambda^2)}^2 \lambda \eta
\left[
y_4 + \frac{y_7-y_9+y_{10}}{2} +y_6 \frac{Y_6}{X}+{2 \over 3}
y_8 \frac{\tilde{Y}_8-\frac{3}{4}Y_6}{X}
\right],
 \nn \\
&&Im a_2={-\sqrt{3} G_F \over 2} \frac{X}{i}
 {(A \lambda^2)}^2 \lambda \eta  
\left[
{-y_7+y_9+y_{10}}+\frac{2}{3} y_8 \frac{\tilde{Y}_8}{X}
\right],
\eea
where matrix elements are denoted by $X$, $Y_6$, and ${\tilde Y}_8$.
$X$ is the matrix element of current$\times$current
type operators, $Y_6$ corresponds to the matrix element of
a density$\times$density QCD penguin operator, ${\tilde Y}_8$ is the
matrix element of EW penguin operator. Their derivation and
precise definition will be given in the next section and
are summarized in Table I and II.

\section{Non-linear $\sigma$ model including scalar mesons and
the matrix elements of QCD and EW Penguin operators }
%
The non-linear $\sigma$ model with higher resonances are studied
in \cite{EGPR,EKW}.
In $\Kppg$ decays, in large $N_c$ limit,
scalar meson may contribute
to the matrix elements of
density$\times$density type four-fermi operators ($Q_6$, $Q_8$).
For current$\times$current type four-fermi interactions,
the amplitude is proportional to the form factor of semi-leptonic
decay, i.e., ${f_+}({M_K}^2-{M_{\pi}}^2)+{f_-}M_{\pi}^2$. 
Because the form factors $f_{\pm}(q^2)$ near soft-pion limit
($q^2={M_\pi}^2$) are important, vector mesons contribution to 
the form factors is small and their effect can be safely neglected.
Therefore we include only scalar mesons in chiral lagrangian,
\bea
{\cal L}&=&{f^2 \over 4} Tr \dU\dUd +B~ Tr {\cal M} (U+U^\dagger)
         +{\ga \over 4}
          Tr \dU\dU^\dagger{ \xi S \xi^{\dagger}} \nn
             \\
         &+& \gb~ Tr {\cal M}(\xi S \xi + \xi^\dagger S \xi^\dagger )
         + Tr(\DS\DS -{\M}^2 S^2),
\label{eq:lagrangian}
\eea
where ${\cal M} = diag(m_u,m_d,m_s)$, $U = \exp\left(i 2 \pi / f\right)=\xi^2$ and $S$
is a scalar nonet field,
\bea
S = \frac{1}{2}
\left(
\begin{array}{ccc}
       \sigma+\delta^0 & \sqrt{2} \delta^0 & \sqrt{2} \kappa^+\\ 
       \sqrt{2}\delta^0 & \sigma-\delta^0 & \sqrt{2} \kappa^0\\ 
       \sqrt{2} {\kappa^-} & \sqrt{2} \bar{\kappa^0}& \sqrt{2} \delta_{ss} 
\end{array} 
\right).
\eea
$\DS$ is covariant derivative and is defined by,
\bea
D_{\mu} S&=& \partial_{\mu} S + i[\alpha_{// \mu}, S] ,\\
\alpha_{// \mu}&=&{\xi^{\dagger} \partial_{\mu} \xi
+ \xi \partial_{\mu} \xi^{\dagger} \over 2i}
= {[\pi, \partial_{\mu} \pi] \over 2 i f^2 }+ \ldots 
\eea
In the lagrangian, the scalar mesons couple to pions through two
terms denoted by $g_1$ and $g_2$.
One is a coupling in SU(3) limit and the other is a coupling with
SU(3) breaking.
The mass splitting term for the scalar nonets
and isospin breaking effect are neglected.
By shifting the scalar meson fields from their vacuum expectation value,
\bea
&& S \rightarrow S+<S>,  ~~~<S>={\gb \over {\M}^2} {\cal M},
\eea
we obtain the mass formulae and decay constants \cite{EGPR}.
They are given in the appendix B.
The parameters in the chiral lagrangian can be written in terms of
physical quantity $F_K, F_\pi, M_K, M_\pi$ and quark masses $m_u(=m_d)$ and
$m_s$.
\bea
&& B={2 \over m_s (1-\Delta)}
    \left[ {(\Delta+1) \Mp^2 \Fp^2 \over 8 \Delta}
          -{\Delta \Mk^2 \Fk^2 \over 2 (1+\Delta)}
   \right],
\\
&& {\gb^2 \over {\M}^2}={2 \over m_s^2 (1-\Delta)}
    \left[-{\Fp^2 \Mp^2 \over 4 \Delta}
          +{\Fk^2 \Mk^2 \over 2 (1+\Delta) }
    \right],
\\
&& {\ga \gb \over {\M}^2}={2 (\Fk^2-\Fp^2) \over m_s (1-\Delta)},
\eea
where $\Delta={m_u / m_s}$.
For computation of the weak matrix elements,
we need  strong interaction vertices.
They can be found in appendix C.

In our calculation using the lagrangian Eq.(\ref{eq:lagrangian}),
a certain class of the higher order terms in the CHPT are summed up due 
to the effect of the scalar resonance exchange.   
These are very important in the process $K\rightarrow \pi \pi$ if the $\sigma$
meson mass is light as kaon mass, $M_\sigma \sim M_K$.
Though systematic treatment of momentum expansion in the CHPT is lost, a class
of the higher order terms in the CHPT are automatically summed up using lagrangian
of Eq.(\ref{eq:lagrangian}).  

Now we turn to the matrix element of QCD and EW penguins.
The explicit derivation is given for two density$\times$density 
type operators $Q_6$ and $Q_8$. Their definitions are,
\bea \displaystyle
Q_6 &=&-8 \sum_{q=u,d,s}({\bar s_L} q_R) ({\bar q_R} d_L), \\
Q_8&=&-12 \sum_{q=u,d,s}({\bar s_L} q_R) e_q ({\bar q_R} d_L)
   = {\tilde Q}_8-\frac{1}{2} Q_6,\\
{\tilde Q}_8&=&-8 ({\bar s_L} u_R) ({\bar u_R} d_L),
\eea
where the subsidiary operator ${\tilde Q}_8$ is introduced.
These operators can be written in terms of meson fields
by identifying quark bilinear as corresponding density,
\bea
\bar{q}_R^j q_L^i=-B U_{ij}-g_2
\left(~\xi ( S + \langle S\rangle) \xi ~ \right)_{ij}.
\eea
After some algebra, we express $Q_6$ in terms of the
meson fields,
\bea
Q_6 &=& -8
\left[~~ \frac{i}{\sqrt{2}} {\gb^2(m_s-m_u) \over  {\M}^2}
         \left(2B + {\gb^2 (m_s+m_u) \over {\M}^2} \right) {K^0 \over \Fk}
\right. \nn \\
&-& \left.
        {i \over 24 \sqrt{2}}{\gb^2 (m_s-m_u)\over {\M}^2}
        \left(2B +  {\gb^2 (m_s+m_u) \over {\M}^2}\right)
         {{K^0 {\pi^0}^2} \over {\Fk \Fp^2}}
\right. \nn \\
&-& \left. {i \gb \over 2 \sqrt{2}}
   \left\{\left(2B+ \gb^2 {(m_s+m_u) \over {\M}^2}\right)
           {\pi^0 \kappa \over \Fp}
      + \left(2B + {\gb^2 2 m_u \over {\M}^2}\right) {\sigma K^0 \over \Fk}
\right\} ~~
\right].
\eea
There are four diagrams which may contribute to $K^0 \rightarrow
\pi^0 \pi^0$ amplitude $Y_6$. (See Feynman diagrams in Fig.3 - Fig.5)
They are classified as follows.\\
1) The diagram in which  $K^0$ decays into $ K^0 \pi^0 \pi^0 $
through the strong vertex
and subsequently $K^0$ vanishes into vacuum through the $Q_6$ (
$T_{tadpole}$).\\
2) The diagram in which $K^0$ directly decays into $2\pi^0$ ($T_{direct}$).
\\
3) The diagram in which $K^0$ is converted into $\sigma$ and subsequently
$\sigma$ decays into
$2 \pi^0$ ($T_{\sigma \hyf pole}$).\\
4) The diagram in which K decays into $ \kappa, \pi^0$
through strong vertex
and $\kappa$ is converted into  $\pi^0$ ($T_{\kappa \hyf pole}$).\\
\def\delpi{\delta_{\pi}}
\def\delk{\delta_{k}}
The sum of the  contribution is denoted by $Y_6$ and it
can be simplified as,
\bea
Y_6 
&=&
{i \sqrt{2} \over \Fk \Fp^2}
\left[ ~
-{\ga \gb \over {\M}^2}
\left\{\Bzb {\Mk^2 \over 1-\delpi^2}+\Bb {\Mk^2-2\Mp^2 \over 1-\delk^2}
\right\}
\right.
\nn \\
&& +{\gb^2 \over {\M}^2}
\left\{ {4 \Bzb \over 1-\delpi^2}
         \left(m_s+m_u-{(m_s-m_u) \Mk^2 \over 2 {\M}^2}\right)
-{8 \Bb m_u \over 1-\delk^2}
-(m_s-m_u) \left(2\Bzb+\Bb(1+{1\over R^2})\right)
\right\} \nn \\
&& -\left.
{\ga \gb^3 \Bb(m_s-m_u)^2 \over 2 \Fk^2 {\M}^4}
+{\gb^4  (m_s-m_u)^2 \over  {\M}^4}
~\right],
\label{eq:y6}
\eea
where $R={\Fk / \Fp}$  and
$\delta_{K (\pi)}={M_{K (\pi)} / \M}$.
We also introduce the auxiliary quantities,
\bea
&&
\Bb=B+{\gb^2 (m_s+m_u) \over 2 {\M}^2}, ~~~~~
\Bzb=B+{\gb^2 m_u \over {\M}^2}.
\eea
They can be written in terms of physical quantities,
\bea
&& 
\Bb={\Mk^2 \Fk^2 \over 2 m_s (1+\Delta) }, ~~~~~~~~
\Bzb={\Mp^2 \Fp^2 \over 4 m_s \Delta }.
\eea
The matrix element of EW penguin operator $Q_8$ is straightforward.
Technically we split $Q_6$ from $Q_8$ so that we do not have to repeat 
the calculation of $Q_6$.  The rest is called ${\tilde Q}_8$ and
given by,
\bea
&& {\tilde Q}_8 =
\left[ i g_2 \Bzb \frac{\pi^- \kappa^+}{F_\pi}
      -i\frac{g_2 \Bzb (\Bzb +\Bb)}{\sqrt{2}}
         \frac{K^0 \pi^+ \pi^-}{F_K F_\pi^2}
\right] + \cdots.
\eea
In $K^0 \rightarrow \pi^+ \pi^-$ amplitude,
there are two contributions to the hadron matrix element
of $\widetilde{Q}_8$, i.e.,
(a) $\kappa$-pole contribution and (b) direct contribution.
The sum is called ${\tilde Y}_8$ and is given by,
\bea
{\tilde Y}_8&=&
  - i \frac{3}{\sqrt{2} F_K F_\pi^2} \frac{\bar{B}_0}{{\M}^2-M_\pi^2}
\left[ 
     g_1 g_2 M_K^2 - 2 g_2^2
     \left\{ 
       (m_s + 3\bar{m}) - (m_s-\bar{m}) 
      \frac{M_K^2-M_\pi^2 }{{\M}^2}
     \right\}
  \right] 
\nonumber\\
&&+  
i 6
\sqrt{2} \frac{1}{F_K F_\pi^2} \bar{B}_0
  \left( \bar{B} + \bar{B}_0 \right).
\eea
Keeping leading terms of $1/M_\sigma$ the matrix elements 
$Y_6, \tilde{Y}_8$ reduce to the well known results $Y_6^0, \tilde{Y}_8^0$
\cite{CHIV}, which correspond to those in the leading order of momentum 
expansion and in large $N_c$ limit.

\begin{eqnarray}
&& 
Y_6^0 = - 4 \sqrt{2} i \left( F_K - F_{\pi} \right) 
            \left( \frac{M_K^2}{m_s+m_u}\right)^2, ~~~~~
\tilde{Y}_8^0 =  3 \sqrt{2} i F_{\pi} 
                     \left( \frac{M_K^2}{m_s+m_u}\right)^2. 
\end{eqnarray}
This approximation is valid only when $M_K \ll M_\sigma$.
In the next section we will show how the values of the matrix
elements of the density$\times$density operators are different 
from $Y_6^0, \tilde{Y}_8^0$ numerically. 

The matrix elements of the other current$\times$current operators
$Q_i (i\ne 6,8)$ are also shown in Table I and II, and 
are expressed by a single amplitude X:
\bea
X=i \sqrt{2} f ({M_K}^2-{M_{\pi}}^2).
\eea

\section{Numerical Results}
In this section, we first estimate the hadronic parameters 
$B_{6, 8}$ corresponding to the matrix element $Q_6$ and $Q_8$
in the factorization approximation.
We compare our results with those from the linear
$\sigma$ model. As an application, we also compute
$\epsilon^{\prime}/\epsilon$ and $Re a_0,Re a_2$.
This is done in isospin limit.

The conventional bag-factors $B_6^{1/2}$, $B_8^{3/2}$, which
people often refer, is defined by the following equation
in our notation,
\begin{eqnarray}
&&
B_6^{1/2}  =  \frac{Y_6}{Y_6^0}, ~~~~~
B_8^{3/2}  =  \frac{2 \tilde{Y_8} - X}{2 \tilde{Y_8^0} - X}.
\end{eqnarray}

As explained in section III,
using $M_{\pi}$, $M_K$, $F_{\pi}$, $F_K$ as inputs
our model can be described by three free parameters in the lagrangian
$M_{\sigma}$, $m_s$, $\Delta = m_u/m_s$ and another parameter $\mu$
in the matching process which is  the factorization scale.

We find that our model predicts that the factorizable part of 
$B_6^{1/2}$ ranges around $1.6 \sim 3.0$ depending on the 
$\sigma$ meson mass $M_{\sigma}$.
The quark mass dependence is not significant.
On the other hand, $B_8^{3/2}$ ranges around $0.7 \sim 1.1$ 
depending on $\Delta$. We varied $\Delta$ in the range of 
$1/20-1/30$ and smaller $\Delta$ gives larger value of $B_8^{3/2}$. 
$M_{\sigma}$ dependence is negligible for $B_8^{3/2}$. 
The numerical value is given in Table III.

Let us now compare our results with those from linear $\sigma$ model
\cite{MLS}. The factorizable bag-factors are given by the following equations.
\begin{eqnarray}
B_6^{1/2} & = & \frac{F_K}{3 F_{\pi} - 2 F_K} \sim 2,  \\
B_8^{3/2} & \simeq & \frac{F_K}{F_{\pi}} \sim 1.2,
\end{eqnarray}
Because there are only four parameters in the linear $\sigma$ model
lagrangian, after using $M_{\pi,K},F_{\pi,K}$ there are no free
parameters left, so that the model predicts $\Delta=1/30$
and $M_{\sigma} \sim 0.9$ GeV.

From Table III we find that $B_8^{3/2}$ from our model with
$\Delta=1/30$ and that from the linear $\sigma$ model are consistent.
It can be seen that the factorizable part of 
$B_6^{1/2}$ for $M_{\sigma}=0.9$ GeV is around 1.5,
which is smaller than the linear $\sigma$ model result.

Next we apply our result to $\epsilon^{\prime}/\epsilon$ and
$Re a_0, Re a_2$. For numerical computation of $\epe$, we use
the experimental values for $Re a_I$.

We have calculated the
next to leading order Wilson coefficients in NDR scheme.
we  could reproduce the numerical values tabulated in Ref. \cite{BBL}
to a good extent.
We chose the following values for the computation,
  $m_t$               = 165.00 GeV,
  $m_W$               =  80.20 GeV,
  $m_b$               =   4.40 GeV,
  $m_c$               =   1.30 GeV,
  $1/\alpha_{QED}$       =   129.0,
  $sin^2 (\theta_W)$  =  0.230,
  $\Lambda_{QCD}^{(5)}$    =  0.226 GeV,
  $\Lambda_{QCD}^{(4)}$    =  0.325 GeV,
  $ \alpha^{\overline{MS}}(m_Z)^{(5)}$   =  0.11799.
We list the Wilson coefficients $z_i,y_i$ at scales $\mu$ = 1.2, 1.0, 0.8 GeV
in Table IV. In this calculation, we used the anomalous dimensions at the NLO 
by Buras et al.

As was explained in section II, in the leading order in large $N_c$ expansion,
$Rea_I, Ima_I$ are obtained by multiplying Wilson coefficients
$z_i(\mu),y_i(\mu)$ with the matrix elements of
$Q_1(\mu), \cdots , Q_{10}(\mu)$ in our model, where $\mu$ is the
factorization scale which is assumed to be 0.8 $-$ 1.2 GeV.

Here we should remark one point about the quark mass.
In large $N_c$ limit, we approximate the matrix elements with
$Q_6(\mu)$ operator by the product of matrix elements
with scalar quark operator at scale $\mu$. Using PCAC
relation, we then convert them to $F_K M_K^2/m_s(\mu)$.
Here, the scale of the strange quark mass should be the same scale
$\mu$. Therefore, when we substitute the mass parameter $m_s$
in our final result, we should run the quark mass to the factorization
scale $\mu$=0.8,1.0,1.2 GeV. For example, $m_s$(2GeV)= 80 $-$ 120 MeV
corresponds to $m_s$(0.8GeV)=136 $-$ 204 MeV. 

\begin{figure}[h]
\begin{center}
\leavevmode 
\psfig{file=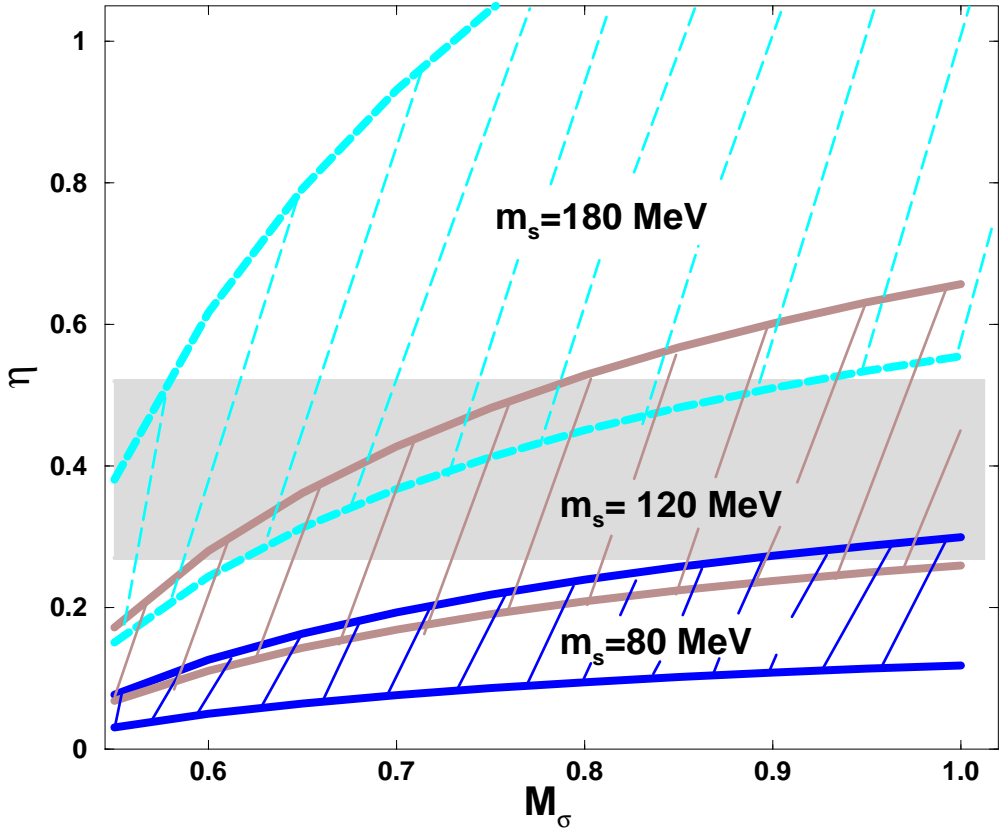,width=16cm,angle=-90}
\end{center}
\caption{Allowed regions for  $M_\sigma$-$\eta$ from $\epsilon'/\epsilon$ 
data at 2-$\sigma$ confidence level. They are shown for
three different values for the strange quark mass, 
i.e.,$m_s$(2GeV)= 80, 120, 180 MeV.}
\label{eta_M}
\end{figure}

Figure \ref{eta_M} shows the dependence of $\eta$ from our model
on the scalar resonance mass $M_{\sigma}$. 
Here, we take $m_s^{\overline{MS}}$(2GeV)=80,120,180 MeV, which 
cover the recent QCD calculations \cite{SUMR,ONO}, and the scale $\mu$ is
chosen to be 0.8 GeV. Upper and lower lines correspond to the maximum 
and minimum values of $\epsilon'/\epsilon$, respectively. 
We find that when $M_{\sigma}$ is larger than $0.8$ GeV 
in order for $\eta$ to lie within  0.27-0.52, which is favored by other 
measurements of CKM parameters, $m_s^{\overline{MS}}$(2GeV) should take 
rather small value 0.09 $-$ 0.12 GeV.
These values are consistent with recent lattice QCD calculations \cite{ONO} 
but smaller compared with QCD sum rule results \cite{SUMR}. 
On the other hand,  as $M_{\sigma}$ becomes smaller
the $Q_6$ amplitude gets enhanced, in order for $\eta$ to lie within  
0.27-0.52, larger value of $m_s^{\overline{MS}}$(2GeV) is preferred.

\begin{figure}[H]
\begin{center}
\leavevmode
\psfig{file=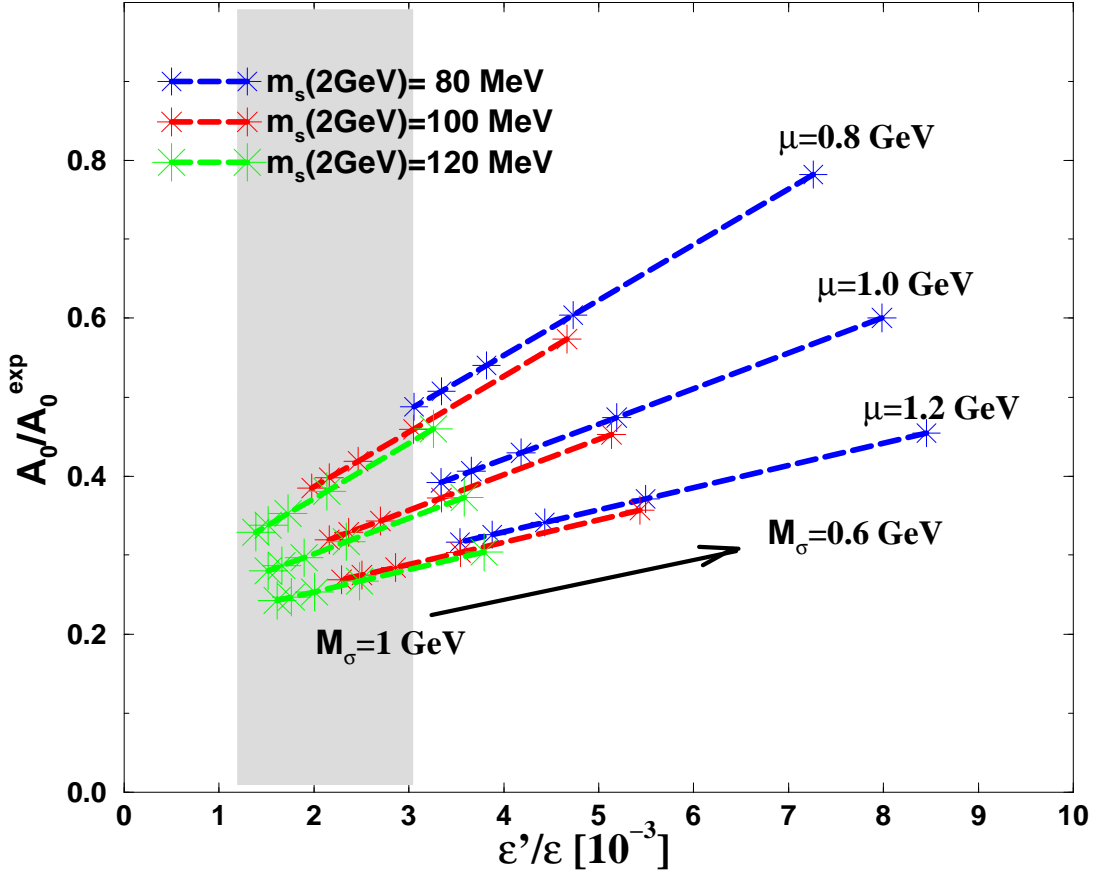,width=17cm,angle=-90}
\end{center}
\caption{Correlation of $a_0$ and $\epsilon^{\prime}/\epsilon$.
The stars on the lines correspond to 
$M_\sigma$ =0.6, 0.7, 0.8, 0.9, 1.0 GeV, $m_s$(2GeV)=80, 100, 120 MeV 
at factrization scale $\mu$ =0.8, 1.0, 1.2 GeV.}
\label{a0}
\end{figure}

Finally, in Figure \ref{a0}, we show the correlation of
$\displaystyle{a_0/ a_0^{\rm exp}}$ and
$\displaystyle{\epsilon^{\prime} / \epsilon}$ by changing the scalar
meson mass from 0.6 GeV to 1 GeV. We take three different values
of $m_s^{\overline{\rm MS}}$(2GeV), which are 80, 100 and 120 MeV.
We take three values for the factorization scale $\mu$,
which are 0.8, 1.0 and 1.2 GeV. We used the experimental
values of $Re a_I$ for the numerical analysis of $\epe$.
CP violation parameter is chosen to be 0.3 in the figure.
The shaded region is the experimental data from KTeV and NA48 
for $\displaystyle{\epsilon^{\prime}/ \epsilon}$ at 2-$\sigma$ confidence level.

We find that $\displaystyle{\epsilon^{\prime} / \epsilon}$
can be easily explained in our model by a suitable choice of the parameters.
Typical value of $M_{\sigma}$, and $m_s^{\overline{\rm MS}}$(2GeV)
is around 0.8 GeV and around 80 MeV respectively almost independent
of the factorization scale $\mu$.
On the other hand, $Rea_0$ in our model are smaller than experiment.
We find that it is quite sensitive to the factorization scale $\mu$,
and as $\mu$ gets smaller, $Rea_0$ becomes larger towards the
experimental  value. For $\mu$=0.8 GeV,
$\displaystyle{a_0 / a_0^{\rm exp}}$ is around 0.5 $-$ 0.6.

The sensitivity of $a_0$ amplitude on $\mu$ can be understood as follows.
The Wilson coefficient $z_6(\mu)$ vanishes when the GIM cancellation
between the charm penguin and up penguin loop is exact. In our
calculation, since we take the $\overline{MS}$ scheme, the cancellation
is exact above charm threshold. Therefore, $z_6(\mu)$ takes nonzero
value only when $\mu < m_c$. Since the factorization scale is very close
to the charm threshold, the result of $z_6(\mu)$ changes quite a lot.

In contrast, the Wilson coefficient $y_6(\mu)$
vanishes when the GIM cancellation between the top penguin charm penguin
loop is exact. Since top decouples
already below $M_W$, this cancellation is completely violated and
$y_6(\mu)$ takes nonzero value from the start and keeps growing all the
way down to the factorization scale. Since $log(M_W/1.2 \mbox{GeV})$ and
$log(M_W/0.8 \mbox{GeV})$ is almost identical, $y_6(\mu)$ is not so sensitive
to the factorization scale.
About $Re a_2$, our result is about $1.5$ times larger than the experimental 
value.
\section{Summary and Discussion}
In this paper, we study the correlation of $\Delta I=1/2$ amplitude
and $\epe$ in the framework of non-linear $\sigma$ model including
the scalar mesons.
We have calculated the matrix elements of the QCD and EW penguins using that
model and within the factorization approximation.

We can not find the scalar meson mass region which
is compatible with both $\epe$ and $\Delta I=1/2$ amplitude simultaneously. 
The reason is follows. We can read from Fig.2, the maximum allowed
value for $(\epe)/\eta$ is about $0.01$.
The bag-factor ${B_6}^{1/2}$ required for $\epe$ is
at most $2 - 3$, which corresponds to
$M_\sigma|_{min} \simeq 0.6 - 0.7$ GeV.
In the range of the scalar meson mass, about $half$ of $\Delta I=1/2$ amplitude
may be explained. Therefore, if we impose the $\epe$ constraint,
we can not explain the whole $\Delta I=1/2$ amplitude.
Moreover, $\epe$ is rather stable for the change
of the factorization scale. This suggests that the prediction of
$\epe$ may be more reliable.
Though there is strong correlation between $\Delta I=1/2$ amplitude and $\epe$,
we conclude the understanding of $\Delta I=1/2$ rule may not be complete. 

Finally, we argue what kind of effects may remedy the
problem. Because QCD penguin $Q_6$ is born just below $m_c$,
the coefficient is not stable about the change of the
factorization scale around 1 GeV.
In the scheme, in which GIM cancellation is incomplete above
$\mu \ge m_c$, the leading order results of the Wilson coefficient 
of $Q_6$ becomes larger by a factor of 2 \cite{BBG}.
This effect was not incorporated in the Wilson coefficients of NLL 
approximation employed here. 
Therefore the same effect may further enhance the Wilson 
coefficient of $Q_6$ used in our analysis.
We also note that the real part of the $\Delta I=3/2$ amplitude
is larger by a factor of 1.5 than the experimental value.
This may tell us  that there is some
suppression (enhancement) coming from the low energy evolution
(pion loops) for CP conserving $\Delta I=3/2$ ($\Delta I=1/2$) amplitudes 
\cite{BBG2,BURAS,BEFL,HKS,HKPSB}. 
A plausible explanation is that the non-factorizable contributions
are very large.
Including these effects may help for
the entire understanding  of both $\epe$ and $\Delta I=1/2$ rule.

\vspace{1cm}
{\bf Acknowledgements} \\
We would like to thank T. Yamanaka, C. S. Lim and U. Nierste for fruitful
discussion and comments.
Y.-Y. K. is grateful to M. Kobayashi
for his encouragement. He would like to thank
 C.D. Lu for their hospitality during
his staying at Hiroshima University.
His work is supported by the Grant-in Aid for Scientific
from the Ministry of Education, Science and Culture, Japan.
Work of T. M. is supported by the Grant-in Aid for Scientific
Research (Physics of CP violation)
from the Ministry of Education, Science and Culture, Japan.

\newpage
\appendix

\section{\bf $\Delta I =1/2$ rule and $\epsilon'/\epsilon$ }
Here we summarize isospin amplitudes and $\epsilon'/\epsilon$.
\bea
\Gamma[\Kpipi]&=&Br(\Kpipi) \times 1/\tau_s,\\
\Gamma[\Kpm]&=&Br(\Kpm) \times 1/\tau_s, \\
\Gamma[\Kpp]&=&Br(\Kpp) \times 1/\tau_+,
\eea
where
\bea
&&\tau_s=(0.8927 \pm 0.0009) 10^{-10} sec,\\
&&\tau_+=(1.2386 \pm 0.0024) 10^{-8} sec,\\
&&{1 \over \tau_+}=5.3142 \times 10^{-14} \mbox{MeV},\\
&&Br(\Kpipi)=31.39 \pm 0.28\% ,\\
&&Br(\Kpm)=68.61 \pm 0.28 \% ,\\
&&Br(\Kpp)=21.16 \pm 0.14 \% ,
\eea
\bea
|I=0\rangle &=& \sqrt{1 \over3} |\pi^0 \pi^0\rangle 
              + \sqrt{2 \over 3} |\pi^+ \pi^-\rangle ,\\
|I=2, I_3=0\rangle &=& 
               -\sqrt{2 \over3} |\pi^0 \pi^0\rangle 
               +\sqrt{1 \over 3} |\pi^+ \pi^-\rangle ,\\
|I=2, I_3=1\rangle &=& |\pi^0 \pi^+\rangle ,
\eea
where $|\pi^+ \pi^-\rangle$ and $| \pi^0 \pi^+\rangle$ are the 
symmetrized states defined as 
\bea
|\pi^+ \pi^- \rangle &=& (|\pi^+\rangle \times |\pi^-\rangle 
                         +|\pi^-\rangle \times |\pi^+ \rangle)/\sqrt{2},\\
|\pi^+ \pi^0 \rangle &=& (|\pi^+\rangle \times |\pi^0\rangle
                         +|\pi^0\rangle \times |\pi^+\rangle)/\sqrt{2}.
\eea
We can write the decay rates in terms of the isospin amplitudes:
\bea
\langle \pi^0 \pi^0|H_w|K_s \rangle 
&=& 
i\sqrt{2} \{\sqrt{{1 \over 3}} Re a_0 \exp{i\delta_0}-\sqrt{{2 \over 3}} 
Re a_2 \exp{i \delta_2} \},
\\
\langle \pi^+ \pi^-|H_w|K_s \rangle 
&=&
i \sqrt{2} \{\sqrt{{2 \over 3}} Re a_0 \exp{i\delta_0}
+\sqrt{{1 \over 3}} Re a_2 \exp{i \delta_2} \},
\\
\langle \pi^+ \pi^0|H_w|K^+ \rangle 
&=& 
i \sqrt{{3 \over 2}} a_2,
\eea
where $i a_I \exp(i \delta_I) =\langle I|H_w|K^0\rangle, I=0,2$ and
$|K_s \rangle \simeq \frac{1}{\sqrt{2}} (|K_0\rangle-|{\bar K_0}\rangle).$
With the definition, we can write:
\bea
\Gamma({\Kpipi})&=&P{2 \over 3}| Re a_0 \exp{i \delta_0}-\sqrt{2} Re a_2
\exp{i \delta_2}|^2 {1 \over 2},\\
\Gamma({\Kpm})&=&P{2 \over 3}| \sqrt{2} Re a_0 \exp{i \delta_0}+ Re a_2
\exp{i \delta_2}|^2 {1 \over 2},\\
\Gamma({\Kpp})&=& P{ 3\over 4} {Re a_0}^2,
\eea
where imaginary parts are neglected.
$P$ is a phase space factor of two body decay and is defined as:
\bea
P&=& {1 \over 16 \pi M_K} \sqrt{1- 4 M_{\pi}^2/{M_k}^2}\\
&=& 3.34919 \times 10^{-5} (MeV^{^1}).
\eea
Here we use $\quad M_{K^+}=493.677$ MeV and 
${\bar M}_{\pi}=({M_{\pi^0}+M_{\pi^+}})/2=137.273$ MeV.
With these definitions, we obtain,
\bea
{Br(\Kpp) \over {Br(\Kpipi)+Br(\Kpm)}}{\tau_s \over \tau_+}=
{3 \over 4} {{Re a_2}^2 \over {Re a_2}^2 +{Re a_0}^2}.
\eea
We can extract the following ratio and values for  $a_0$ and $a_2$,
\bea
{Re a_0 \over Re a_2}&=&22.15,\\
Re a_2&=&2.114 \times 10^{-5} (\mbox{MeV}),\\
Re a_0&=&4.686 \times 10^{-4} (\mbox{MeV}).
\eea

\section{\bf Decay constants, mass formulae}
In this appendix, we collect the formulae for the decay constants
and masses which can be derived using Eq.(6).
\bea
\Mp^2&=& \frac{1}{F_\pi^2} [4 B m_u + 4 {\gb^2 {m_u}^2 \over {\M}^2}],\\
\Mk^2&=& \frac{1}{F_K^2} [2 B (m_s+m_u)  + {\gb^2 (m_s+m_u)^2 \over
{\M}^2}],\\
\Fp &=&\frac{f}{\sqrt{Z_\pi}}[1+{\ga \gb   m_u \over {\M}^2 f^2 }],\\
\Fk &=&\frac{f}{\sqrt{Z_K}}[1+{\ga \gb (m_u + m_s) \over 2 {\M}^2 f^2 }],\\
Z_{\pi}&=&1+{\ga \gb m_u \over {\M}^2 f^2}, \\
Z_{K}&=&1+{\ga \gb (m_u + m_s)  \over 2 {\M}^2 f^2},
\eea
where $Z_{\pi}$ and $Z_K$ are wave function renormalization constants,
$F_{\pi}$ is 92.42 MeV.

\section{\bf  Lagrangian}
Here we record the part of the lagrangian which is relevant 
for calculation.
\bea
{\cal L}_{4 \pi} 
&=& 
- {\pi^2 \dK \dKb \over 12 \Fp^2}
-{\K \Kb \dpi^2 \over  12 \Fk^2}
+ {\pi^0 \dpi ( \K \dKb + \Kb \dK)  \over 24}
\left({1 \over \Fk^2} + {1 \over  \Fp^2}\right) 
\nn \\
&& + {{\pi^0}^2 \K \Kb \over 12 \Fp^2 \Fk^2} 
     \left( \Fp^2 \Mp^2 + \Fk^2 \Mk^2 
           - {3 \gb^2 (m_s-m_u)^2 \over 4 {\M}^2} 
     \right),
\eea
\bea
{\cal L}_{s \pi \pi}
& = &
{\ga \over 4}
\left[ \sigma
\left( 
                     {(\dpi)^2 \over \Fp^2} 
                   + {\dK \dKb \over  \Fk^2}
              \right)
-
\left( {\bar \kappa^0} {\dpi \dK \over \Fp \Fk} + h.c. \right)
\right]
\nn \\
&&
- \gb
\left[ \sigma \left( {(\pi^0)^2 \over \Fp^2 } m_u
                   + {\K \Kb \over \Fk^2}
                     \left({m_u + m_s\over 2}\right)
             \right)
-
  \left(  {\bar \kappa^0}
          {\pi^0 \over \Fp} { \K \over \Fk}
        \left( 3 m_u  + m_s \over 4\right) +h.c.
   \right)
\right]\nn \\
&& ~~+ ~~{\cal \delta L}_{s \pi \pi}~~+~~{\cal \delta L}_{4 \pi},\nn \\
\eea
where ${\cal \delta L}_{s \pi \pi}$ and ${\cal \delta L}_{4 \pi}$
come from the covariant derivative term. (See Eq.(8) and Eq.(9).)
Their explicit forms are, 
\bea
{\cal \delta L}_{s \pi \pi}
&=&
{\gb (m_s-m_u)  \over {\M}^2 4 \Fk \Fp}
\left[ \partial {\bar \kappa^0} ( K\lrpartial\pi^0) + h.c. \right],
\\
{\cal \delta L}_{4 \pi}&=&{\gb^2 (m_s-m_u)^2  \over 16  {\M}^4 \Fk^2 \Fp^2}
\left[(K^0 \lrpartial \pi^0)(\Kb \lrpartial \pi^0 ) \right].
\eea

\section{The matrix element of $Q_6$}
We give the derivation of the matrix element of $Q_6$.
\bea
Y_6&=&T_{tadpole}+T_{direct}+T_{\sigma\hyf pole}+T_{\sigma\hyf tad}+
T_{\kappa\hyf pole}+T_{\kappa \hyf tad}.
\eea
The explicit expression of the parts of Eq.(\ref{eq:y6}) is given by,
\bea
T_{tadpole}+T_{direct}
&=&
-\sqrt{2}i {\gb^2 \over {\M}^2 \Fk}(m_s-m_u) {\bar B}
\nn \\
& \times & 
\left\{ {1 \over \Fk^2}+{ 1 \over  \Fp^2}
- {\gb^2 (m_s-m_u)^2 \over M^4 \Fk^2 \Fp^2}
-{\gb^2 (m_s-m_u)^2 \over \Fk^2 \Fp^2 {\M}^2 \Mk^2}
+{\gb^2 (m_s-m_u)^2 \Mp^2 \over \Fk^2 \Fp^2 {\M}^4 \Mk^2}
\right\},
\nn \\
 T_{\sigma\hyf pole}+T_{\sigma\hyf tad}
&=& 
{i \sqrt{2} \Bb \over \Fk \Fp^2 (\Mk^2-{\M}^2) } 
\left\{ - \ga \gb \left(2\Mp^2-\Mk^2\right)
        +8  \gb^2  m_u  
\right\},
\nn \\
T_{\kappa\hyf pole}+T_{\kappa\hyf tad}
&=&
{i \sqrt{2}\Bb  \over \Fk \Fp^2 ({\M}^2-\Mp^2)}
\left\{
        1-{\gb^2 (m_s-m_u)(m_s+m_u) \over  \Ms^2 \Fk^2 \Mk^2}
       +{ \gb^2 (m_s-m_u)^2 \over  2 \Fk^2 \Mk^2 {\M}^2} 
          \left(1-{\Mp^2 \over {M}^2}\right)
\right\}
\nn \\
& \times & 
\left\{
- \ga \gb \Mk^2 + 4 \gb^2 (m_s+m_u)
- 2 (m_s-m_u) \gb^2 {\Mk^2 \over {\M}^2}-2(m_s-m_u) \gb^2
                                         \left(1- {\Mp^2 \over {M}^2}\right)
\right\}.
\nn \\
\eea



\newpage
\begin{figure}[H]
\begin{center}
\leavevmode
\psfig{file=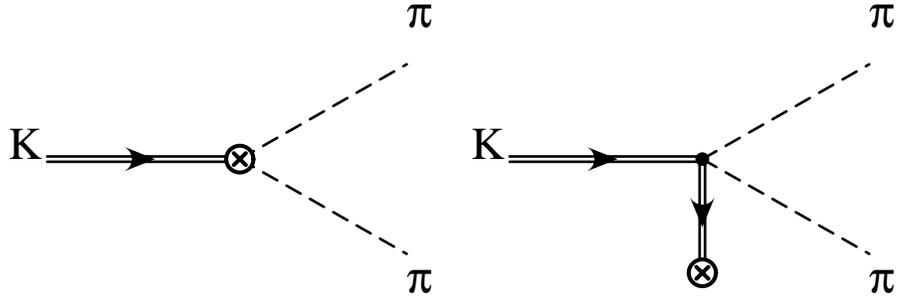,width=12cm}
\end{center}
\caption{Feynman diagrams for $T_{direct}+T_{tadpole}$.}
\label{f1}
\end{figure}

\begin{figure}[H]
\begin{center}
\leavevmode
\psfig{file=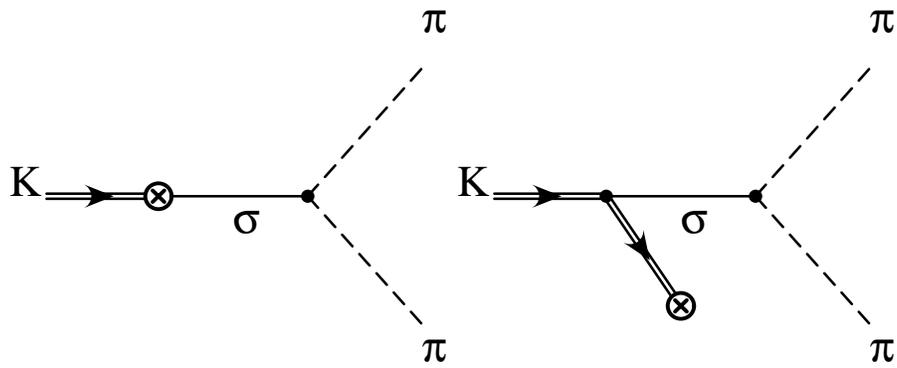,width=12cm}
\end{center}
\caption{Feynman diagrams for $T_{\sigma\hyf pole}+T_{\sigma\hyf tad}$.}
\label{f2}
\end{figure}

\begin{figure}[H]
\begin{center}
\leavevmode
\psfig{file=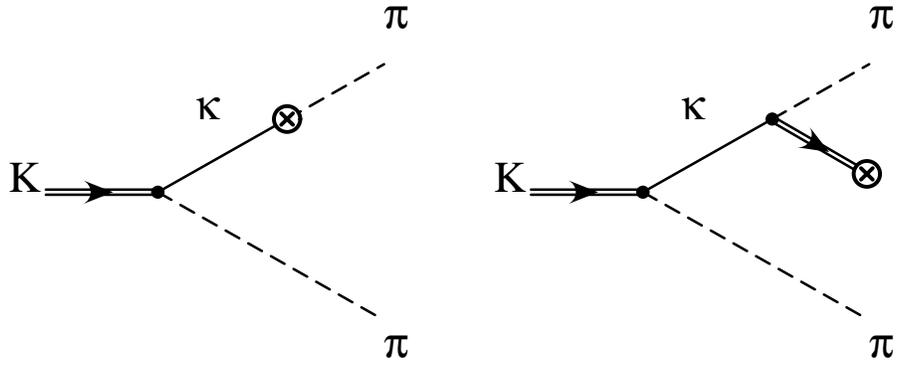,width=12cm}
\end{center}
\caption{ Feynman diagrams for $T_{\kappa\hyf pole}+T_{\kappa\hyf tad}$.}
\label{f2}
\end{figure}

\begin{table}[h]
\begin{center}
\begin{tabular}{cccc} \hline
&{$ \pi^0 \pi^0$} & {$ |\pi^+>\times |\pi^->$}& {$ |\pi^0> \times |\pi^+>$ }
\\ \hline
$Q_1$ & $-X$ &$0$ & $X /\sqrt{2}$ \\
$Q_2$ & $0$ &$X$ & $X /\sqrt{2}$ \\
$Q_3$ & $0$ &$0$ & $0$ \\
$Q_4$ & $X$ &$X$ & $0$ \\
$Q_5$ & $0$ &$0$ & $0$ \\
$Q_6$ & $Y_6$ &$Y_6$ & $0$ \\
$Q_7$ & $3X/2$ &$0$ & $-3 X/2\sqrt{2}$ \\
$\tilde{Q}_8$ & $0$&$\tilde{Y}_8$ & $\tilde{Y}_8/\sqrt{2}$ \\
$Q_9$ & $-3X/2$ &$0$ & $3X/2\sqrt{2}$ \\
$Q_{10}$ & $-X/2$ &$X$ & $X/\sqrt{2}$ \\
$Q_{11}$ & $X$ &$0$ & $X/\sqrt{2}$ \\ \hline
\end{tabular}
\label{tab:matrix}
\caption{The matrix elements, where $X=i \sqrt{2} f ({M_k}^2-{M_{\pi}}^2)$.
$\tilde{Y}_8$ and
$Y_6$ are defined in the text. }
\end{center}
\end{table}

\begin{table}
\begin{center}
\begin{tabular}{ccc} \hline
&{$ a_0$} & {$ a_2$} \\ \hline
  ~~$Q_1$ & ~~$-\sqrt{1\over 3}X$ & ~~$\sqrt{2 \over 3} X$   ~~ \\
  ~~$Q_2$ & ~~${2 \over \sqrt{3}} X $ & ~~$\sqrt{2 \over 3} X$  ~~  \\
  ~~$Q_3$ & ~~$0$ &  ~~$0$    ~~\\
  ~~$Q_4$ & ~~$\sqrt{3}X$ & ~~$0$  ~~  \\
  ~~$Q_5$ & ~~$0$ &  ~~$0$   ~~ \\
  ~~$Q_6$ & ~~$\sqrt{3}Y_6$ &  ~~$0$   ~~\\
  ~~$Q_7$ & ~~${\sqrt{3} \over 2}X$ &  ~~$-\sqrt{3 \over 2} X$   ~~  \\
  ~~$Q_8$ & ~~${2 \over \sqrt{3} } \tilde{Y}_8-{\sqrt{3} \over 2} Y_6$
      &  ~~${\sqrt{2 \over 3}}\tilde{Y}_8$   ~~\\
  ~~$Q_9$ & ~~$-{\sqrt{3} \over 2}X$ &  ~~$\sqrt{ 3\over 2} X$  ~~  \\
  ~~$Q_{10}$ & ~~${\sqrt{3}\over 2} X $ &  ~~ $\sqrt{3 \over 2} X $  ~~\\
\hline
\end{tabular}
\label{tab:isospin}
\caption{Contribution to isospin amplitudes.}
\end{center}
\end{table}

\begin{table}[h]
\begin{center}
\begin{tabular}{|l|l|llllll|} \hline
& $M_\sigma $ (GeV) &  0.55  &  0.6  &  0.7  &  0.8  &  0.9  &  1.0 \\
\cline{2-8}
\hspace{5mm} $\Delta=1/20$ \hspace{5mm} 
& $B^{1/2}_6$    
& 5.27
& 3.29
& 2.22
& 1.84
& 1.64
& 1.52
\\ 
& $B^{3/2}_8$    
&0.70
&0.73
&0.77
&0.79
&0.81
&0.82
\\
\hline
& $M_\sigma$ (GeV) &  0.55  &  0.6  &  0.7  &  0.8  &  0.9  &  1.0  \\
\cline{2-8}
\cline{2-8}
\hspace{5mm} $\Delta=1/25$  \hspace{5mm} 
& $B^{1/2}_6$    
&  4.81 
&  3.06 
&  2.12 
&  1.78 
&  1.61 
&  1.50  \\ 
& $B^{3/2}_8$    
&   0.90 
&  0.93 
&  0.96 
&  0.99 
&  1.00 
&  1.01 \\
\hline
& $M_\sigma$ (GeV) &  0.55  &  0.6  &  0.7  &  0.8  &  0.9  &  1.0  \\
\cline{2-8}
\hspace{5mm} $\Delta=1/30$ \hspace{5mm}
& $B^{1/2}_6$    
& 4.48
& 2.90
& 2.05
& 1.75
& 1.60
& 1.51
\\ 
& $B^{3/2}_8$    
& 1.13
& 1.15
& 1.17
& 1.19
& 1.20
& 1.21
\\
\hline
\end{tabular}
\label{tab:bag}
\end{center}
\caption{Bag-factor}
\end{table}

\begin{table}[h]
\begin{center}
\begin{tabular}{lllll}
\hline
Wilson coeff. & $\mu= 80.2$ GeV & $\mu= 1.2$ GeV & $\mu=1.0$ GeV & $\mu=0.8$ GeV\\
\hline
$y_1$                & 0.0        & 0.0      & 0.0      &  0.0     \\
$y_2$                & 0.0        & 0.0      & 0.0      &  0.0     \\
$y_3$                & 0.0014715  & 0.03058  & 0.03335  &  0.03722 \\
$y_4$                & -0.0019375 & -0.05871 & -0.05884 & -0.05844 \\
$y_5$                & 0.0006458  & 0.00311  & -0.00168 & -0.01384 \\
$y_6$                & -0.0019375 & -0.09797 & -0.11672 & -0.16226 \\
$y_7/\alpha_{QED}$    & 0.1262367  & -0.03714 & -0.03822 & -0.04038 \\
$y_8/\alpha_{QED}$    & 0.0        & 0.14352  & 0.17174  &  0.23136 \\
$y_9/\alpha_{QED}$    & -1.0606455 & -1.46549 & -1.54058 & -1.69377 \\
$y_{10}/\alpha_{QED}$ & 0.9        & 0.57829  & 0.68795  &  0.89882 \\
$z_1$                & 0.0526643  & -0.45108 & -0.52381 & -0.64505 \\
$z_2$                & 0.9812457  & 1.23913  & 1.28816  &  1.37464 \\
$z_3$                & 0.0        & 0.00674  & 0.01353  &  0.03059 \\
$z_4$                & 0.0        & -0.01980 & -0.03704 & -0.07439 \\
$z_5$                & 0.0        & 0.00569  & 0.00784  &  0.00844 \\
$z_6$                & 0.0        & -0.01950  & -0.03698 & -0.08023 \\
$z_7/\alpha_{QED}$    & 0.0        & 0.00940  & 0.01249  &  0.01989 \\
$z_8/\alpha_{QED}$    & 0.0        & 0.00349  & 0.01551  &  0.04725 \\
$z_9/\alpha_{QED}$    & 0.0        & 0.01127  & 0.02019  &  0.03993 \\
$z_{10}/\alpha_{QED}$ & 0.0        & -0.00219  & -0.00893 & -0.02287 \\
\hline
\end{tabular}
\label{tab:wilson_coeff}
\end{center}
\caption{List of Wilson coefficients $y_i,z_i$.}
\end{table}
%

\end{document}